\documentclass[letter, twocolumn]{jpsj3}
\usepackage{amsmath, amssymb,amsfonts}
\usepackage{stmaryrd}
\usepackage{graphicx}
\usepackage{color}

\usepackage{txfonts}
\DeclareSymbolFont{largesymbols}{OMX}{cmex}{m}{n}

\usepackage{bm}

\newcommand{\BPCB}{$\mathrm{(C_5H_{12}N)_2CuBr_4}$}
\newcommand{\DIMPY}{$\mathrm{(C_7H_{10}N)_2CuBr_4}$}
\newcommand{\sul}{Sul-Cu$_2$Cl$_4$}

\newcommand{\hr}{\mathcal{H}_{\mathrm{rung}}}

\newcommand{\Dt}{\Delta_{\mathrm{t}}}
\newcommand{\Ds}{\Delta_{\mathrm{s}}}

\title{
Electron Spin Resonance in Quasi-One-Dimensional Quantum
Antiferromagnets: 
Relevance of Weak Interchain Interactions
}
\author{Shunsuke C. FURUYA$^1$ and Masahiro SATO$^2$}
\inst{$^1$Department of Quantum Matter Physics, University of Geneva,
24 Quai Ernest-Ansermet 1211 Geneva, Switzerland
$^2$
Department of Physics and Mathematics, Aoyama Gakuin University,
Sagamihara, Kanagawa 229-8558, Japan
}
\date{\today}
\abst{
We discuss universal features on the electron spin resonance (ESR) of 
a temperature-induced Tomonaga-Luttinger liquid  phase in a wide class
of weakly coupled $S=1/2$ antiferromagnetic spin chains such as spin
ladders, spin tubes and three-dimensionally coupled spin chains. 
We show that the ESR linewidth of various coupled chains increases
with lowering temperature while the linewidth of a single spin
chain is typically proportional to temperature. This broadening 
with lowering temperature is attributed to anisotropic interchain 
interactions and has been indeed observed in several kinds of 
three-dimensional (3D) magnets of weakly coupled spin chains 
above the 3D ordering temperature. We demonstrate that
our theory can account for anomalous behaviors of the linewidths in an
$S=1/2$ four-leg spin tube compound Cu$_2$Cl$_4 \cdot $H$_8$C$_4$SO$_2$
(abbreviated to Sul-Cu$_2$Cl$_4$) and a three-dimensionally 
coupled $S=1/2$ spin chain compound CuCl$_2\cdot 2$NC$_5$H$_5$.
}
\kword{ESR, spin chains, spin ladders, spin tubes, Tomonaga-Luttinger
liquid, conformal field theory}
\begin{document}

\maketitle

\textit{Introduction. ---}
The Tomonaga-Luttinger liquid (TLL) is a universal concept that
describes gapless metallic states in one-dimensional (1D) quantum
systems instead of Fermi liquid in higher-dimensional
systems~\cite{Giamarchi_book,Cazalilla_RMP_1d,Imambekov_RMP_1d}. 
Today, various sorts of quasi-1D materials are avilable to study 
1D characteristic features. 
For the realization and control of TLL physics, 
quasi-1D quantum antiferromagnets offer an ideal stage 
accessible by rich experimental techniques. 
For example, a field-induced TLL 
phase~\cite{Chitra_ladder,Bouillot_ladder} under a strong magnetic field
have been investigated in recently synthesized 
spin-ladder compounds \BPCB{} and \DIMPY{} by means of 
thermodynamic measurements~\cite{Ruegg_BPCB, Ninios_DIMPY},
neutron scattering~\cite{Hong_DIMPY,Schmidiger_DIMPY,Schmidiger_DIMPY_spec}, 
nuclear magnetic resoannce 
(NMR)~\cite{Sasaki_DIMPY,Klanjsek_BPCB,Mukhopadhyay_BPCB,Jeong_DIMPY}
and electron spin resonance 
(ESR)~\cite{Fujimoto_DIMPY,Cizmar_ESR_BPCB}.

Among the above experimental techniques,
ESR occupies a unique position thanks to its
high sensitivity to breakdown of the spin rotational symmetry.
For instance, ESR measurements uncovered with high accuracy 
that the
TLL acquires a field-induced gap and multiple kinds of 
bulk~\cite{Oshima_CuBenzoate,Affleck_CuBenzoate,Zvyagin_2004,
Zvyagin_2005,Zvyagin_2012,Ozerov_2013,Umegaki_SG_1,Umegaki_SG_2} and 
edge~\cite{Furuya_BBS} excitations 
when a magnetic field is applied to quantum spin chain compounds 
with staggered Dzyaloshinskii-Moriya (DM) interaction. 
On the theoretical side
the ESR has long proposed important
problems~\cite{KuboTomita,MoriKawasaki,KanamoriTachiki,NagataTazuke,OshikawaAffleck,ElShawish_ESR}. 
For 1D antiferromagnets, a recent prominent development of the ESR
theory was given by Oshikawa and Affleck~\cite{OshikawaAffleck}.
Their theory gave theoretical explanations to the 
resonance frequency and the linewidth of the ESR spectrum in 
quantum spin chain compounds. 

\begin{figure}[t!]
 \centering
 \includegraphics[bb = 0 0 800 200, width=\linewidth]{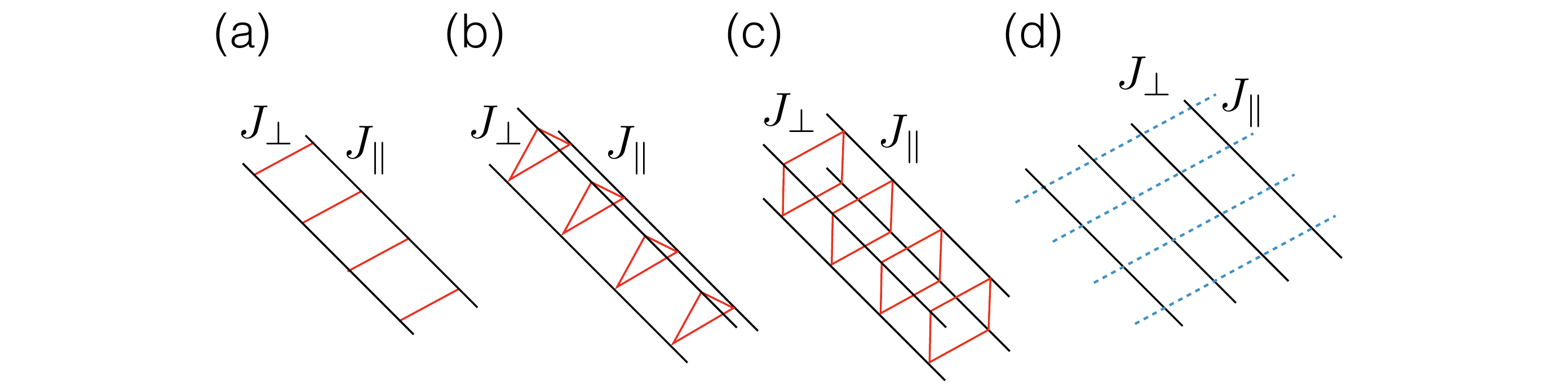}
 \caption{
 (a) Two-leg ladders, (b) three-leg tubes, (c) four-leg tubes,
 and (d) three-dimensionally (3D) coupled spin chains. 
 }
 \label{fig:lattice}
 \centering
 \includegraphics[bb= 0 0 1350 450, width=\linewidth]{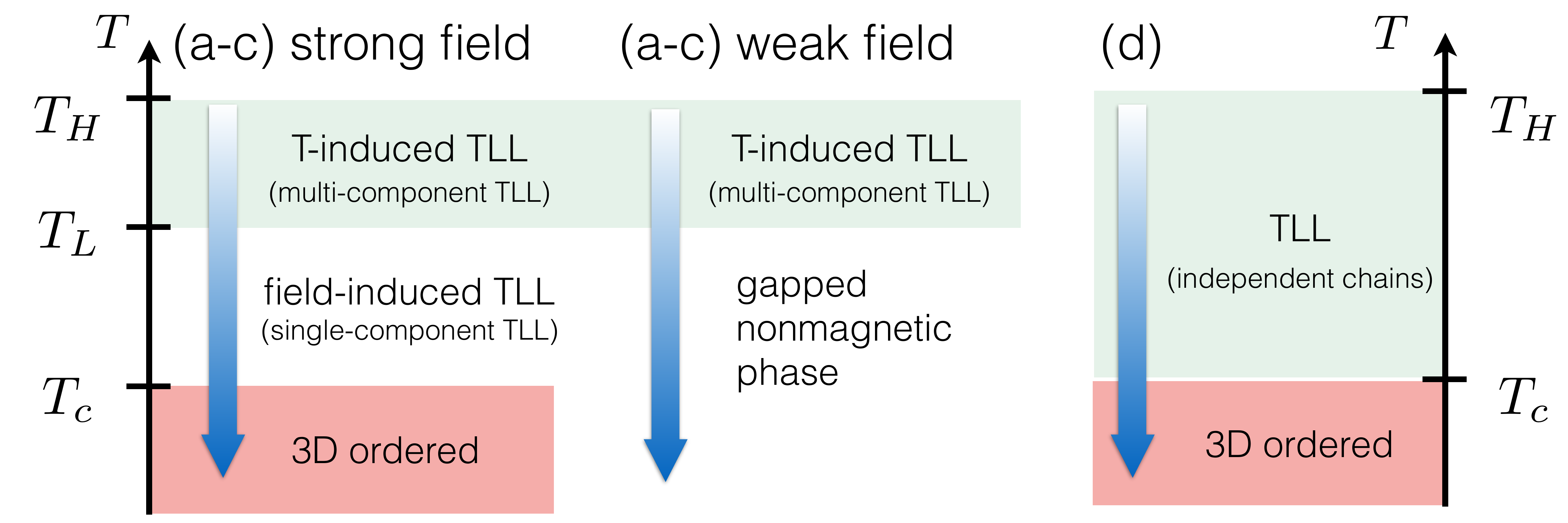}
 \caption{
 Hierarchies of phases in the WCSCs shown in
 Fig.~\ref{fig:lattice}~(a-d).
 The label corresponds to that of Fig.~\ref{fig:lattice}.
 The light green area is the temperature(T)-induced TLL phase of our
 interest. 
 }
 \label{fig:hierarchy}
\end{figure}

One should however notice that we have only a few reliable 
theoretical results~\cite{Furuya_ladder} to understand 
ESR of quasi-1D quantum magnets (Fig.~\ref{fig:lattice}) going beyond the purely 1D 
cases~\cite{Miyashita_ESR_1999,Maeda_XY,Maeda_exact,ElShawish_ESR,Brockmann_ESR,Furuya_NLSM,Furuya_NDMAP}. 
For instance, the ESR linewidth of antiferromagnetically coupled 
quantum spin chains often has a maximum value at 
the N\'eel temperature $T_N$~\cite{Ajiro_ESR}.
In this case, while a phenomenological theory of antiferromagnetic
resonance~\cite{MoriKawasaki} explains the broadening as $T\nearrow
T_N$ (i.e. $T$ approaches $T_N$ from below), no theory explains that as
$T\searrow T_N$ (i.e. $T$ approaches $T_N$ from above).
A quasi-1D antiferromagnet
Cu$_2$Cl$_4 \cdot $H$_8$C$_4$SO$_2$ (abbreviated to
\sul)~\cite{Fujisawa_JPSJ,Fujisawa_PTP,Fujisawa_ESR,Garlea_sul_PRL,
Garlea_sul,Zheludev_sul} is much more interesting.
The ESR linewidth of this compound shows a similar maximum at a
temperature~\cite{Fujisawa_ESR} just like that at $T_N$.
Since this magnet has no magnetic order on the low-temperature 
side~\cite{Fujisawa_JPSJ}, the phenomenological argument is inapplicable. 
This experimental result indicates that the ESR linewidth will be 
capable of probing the crossover to quantum nonmagnetic phases. 
Nevertheless, such behavior of the ESR linewidth is yet to be
understood. 
In order to bridge the gap between the preceding theories and the
potential experimental ability, an organized theory for ESR in 
quasi-1D quantum magnets with interchain interactions is necessary.

In this paper, we present a microscopic theory for universal 
behavior of the ESR linewidth in quasi-1D quantum antiferromagnets,
introducing an idea of the temperature-induced TLL.
For this purpose, we apply field theoretical techniques 
(mainly the bosonization).
We show that the ESR linewidth generally grows near the temperature 
where the TLL phase is violated by interchain interactions. 
Namely, the ESR linewidth provides a good probe to the 
phase transitions or crossovers between temperature-induced TLL and
neighbring lower-temperature phases 
even when the low-temperature phase is nonmagnetic. 
Our theory covers general 
quasi-1D antiferromagnets including 
the aformentioned experiments~\cite{Ajiro_ESR,Fujisawa_ESR}.

\textit{Temperature-induced TLL. ---}
Let us start our discussion from definition of the model for quasi-1D
systems of weakly coupled spin chains (WCSCs) depicted in Fig.~\ref{fig:lattice}. Their Hamiltonian is
given by 
\begin{equation}
 \mathcal{H}=
  \mathcal{H}_\parallel+\mathcal{H}_\perp-g\mu_B H S^z+\mathcal{H}'. 
  \label{eq:H_ladder}
\end{equation}
The first two terms $\mathcal{H}_\parallel$ and $\mathcal{H}_\perp$ 
denote the main part of isotropic intrachain and interchain 
interactions, respectively. 
The third term represents the Zeeman energy with 
the Land\'e factor $g$, the Bohr magneton $\mu_B$, 
and  the external magnetic field $H$ coupled to the total spin
$S^{a}\equiv\sum_{j,n}S^{a}_{j,n}$ ($a=x,y,z$).
The last term $\mathcal{H}'$ is a small anisotropy.
Note that the ESR linewidth must be zero without
$\mathcal{H}'$~\cite{OshikawaAffleck}.

To explain the WCSC \eqref{eq:H_ladder} and the 
temperature-induced TLL, it is instructive to consider a two-leg spin ladder 
with strong exchange interactions along the leg as a simple example 
of WCSCs~\cite{suppl}. The Hamiltonian of the ladder is given by 
\begin{equation}
 \mathcal{H}_\parallel =J_\parallel
  \sum_{n=1,2}\sum_{j}\bm{S}_{j,n}\cdot \bm{S}_{j+1,n}, \quad
  \mathcal{H}_\perp = J_\perp \sum_j \bm{S}_{j,1}\cdot\bm{S}_{j,2},
  \label{eq:H_0}
\end{equation}
where $J_\parallel>0$ is the antiferromagnetic exchange coupling 
along the leg much larger than  the rung coupling $|J_\perp|$ 
(i.e. $J_\parallel\gg|J_\perp|$). 
The anisotropic term $\mathcal{H}'$ is unspecified for the moment.
We will consider several forms of $\mathcal{H}'$ later.
Hereafter we employ a unit $\hbar=k_B=g\mu_B=1$ for simplicity.

The temperature-induced TLL phase is characterized by two energy scales
$T_{L,H}$ (Fig.~\ref{fig:hierarchy}) when external field $H$ 
is sufficiently small. 
The temperature $T_H\sim J_\parallel$ represents an energy cost to flip
an $S=1/2$ spin.
When $T>T_H$, the exchange interaction $J_{\parallel} (\gg |J_{\perp}|)$
is negligible and all the spins fluctuate independently.
In an intermediate region, $T_L<T<T_H$,
the intrachain correlation of spins is developed while
interchain one is \emph{not}. 
The region $T_L<T<T_H$ defines the 
temperature-induced TLL phase where both spin chains
are well described by the TLL and they are only weakly coupled 
through the interchain interaction. 
The lower bound $T_L$ is given by the highest energy $\Delta$ 
in mass gaps of all the magnon bands of WCSCs and therefore 
it depends on the detail of the interchain interaction. 
In the case of the two-leg spin ladder, 
$T_L$ is equal to the excitation gap of the antisymmetric mode, 
which we symbolically denote as 
``$\bm{S}_{j,1}-\bm{S}_{j,2}$''~\cite{suppl,Shelton} and it is 
evaluated as $T_L\sim |J_\perp|$. In the lower temperature regime 
$T<T_L$, the ladder is in a gapped nonmagnetic phase. 
Note that $T_L$ is independent of the external magnetic field $H$ 
because $H$ is coupled to only the symmetric mode 
``$\bm{S}_{j,1}+\bm{S}_{j,2}$'' and not to the antisymmetric mode. 
This nature of $T_L$ is unchanged after 
including $\mathcal{H}'$ such as weak exchange anisotropies. 
By contrast, for example, in the spin chains with a staggered DM interaction, 
$\mathcal{H}'=\sum_{j,n}(-1)^j\bm{D}\cdot\bm{S}_{j,n}\times\bm{S}_{j+1,n}$, 
the excitation gap becomes field dependent and the $T_L$ is evaluated 
as $T_L\propto (DH/J_\parallel^2)^{2/3}$~\cite{Affleck_CuBenzoate}. 
Therefore, the field dependence of $T_L$ allows a clear distinction 
between the staggered DM interaction and the interchain exchange 
interaction.

When the magnetic field becomes strong enough 
in the low temperature regime $T<T_L$, 
the two-leg ladder enters into the field-induced TLL phase 
from the nonmagnetic phase. In the field-induced phase, 
only the symmetric mode ``$\bm{S}_{j,1}+\bm{S}_{j,2}$'' 
survives as the TLL~\cite{Chitra_ladder}. This is the definite 
difference from the temperature-induced TLL. 
When we further lower the temperature, the interladder interaction 
brings about a 3D ordering transition at $T_c$. 
Such a hierarchy of phases is summarized in Fig.~\ref{fig:hierarchy}.

The above argument is easily extended to general 
$N_{\mathrm{leg}}$-leg spin ladders or tubes. 
The upper bound $T_H$ is unchanged and the lower bound $T_L$ is changed 
depending on the form of the interchain interaction. However, $T_L$ 
is still independent of the field. This is because 
only the ``center-of-mass'' mode 
``$\bm{S}_{j,1}+\cdots+\bm{S}_{j,N_{\mathrm{leg}}}$'' is coupled 
to the magnetic field $H$~\cite{OYA} and the field $H$ can make 
the mode gapless, while all of the other ``relative-motion'' modes 
such as ``$\bm{S}_{j,1}-\bm{S}_{j,2}$'' generally 
aquire a finite excitation gap due to the interchain interactions 
($\bm{S}_{j,n}$ denotes spin on $j$ site of the $n$-th chain). 
If the interchain coupling is frustrated, the lower bound 
$T_L/J_\parallel$ will 
be more suppressed than the ratio $J_\perp/J_\parallel$ 
because of (partial) cancelation of neighboring 
frustrated interactions.

Thus the temperature-induced TLL is an $N_{\mathrm{leg}}$-component TLL 
where effects of interchain interactions are weakened by temeprature. 
The temperature $T=T_L$ gives a crossover temperature of the weak 
and the strong interchain interaction regimes.

The case of 3D coupled spin chains [Figs.~\ref{fig:lattice}~(d) and
\ref{fig:hierarchy}~(d)] 
immediately follows by taking $N_{\mathrm{leg}}\to+\infty$ and 
replacing $T_L$ to the 3D ordering temperature 
$T_c\propto|J_\perp|$~\cite{Schulz_CoupledChains, Yasuda_TN}.

\begin{figure}[t!]
 \centering
 \includegraphics[bb=0 0 1200 600,width=\linewidth]{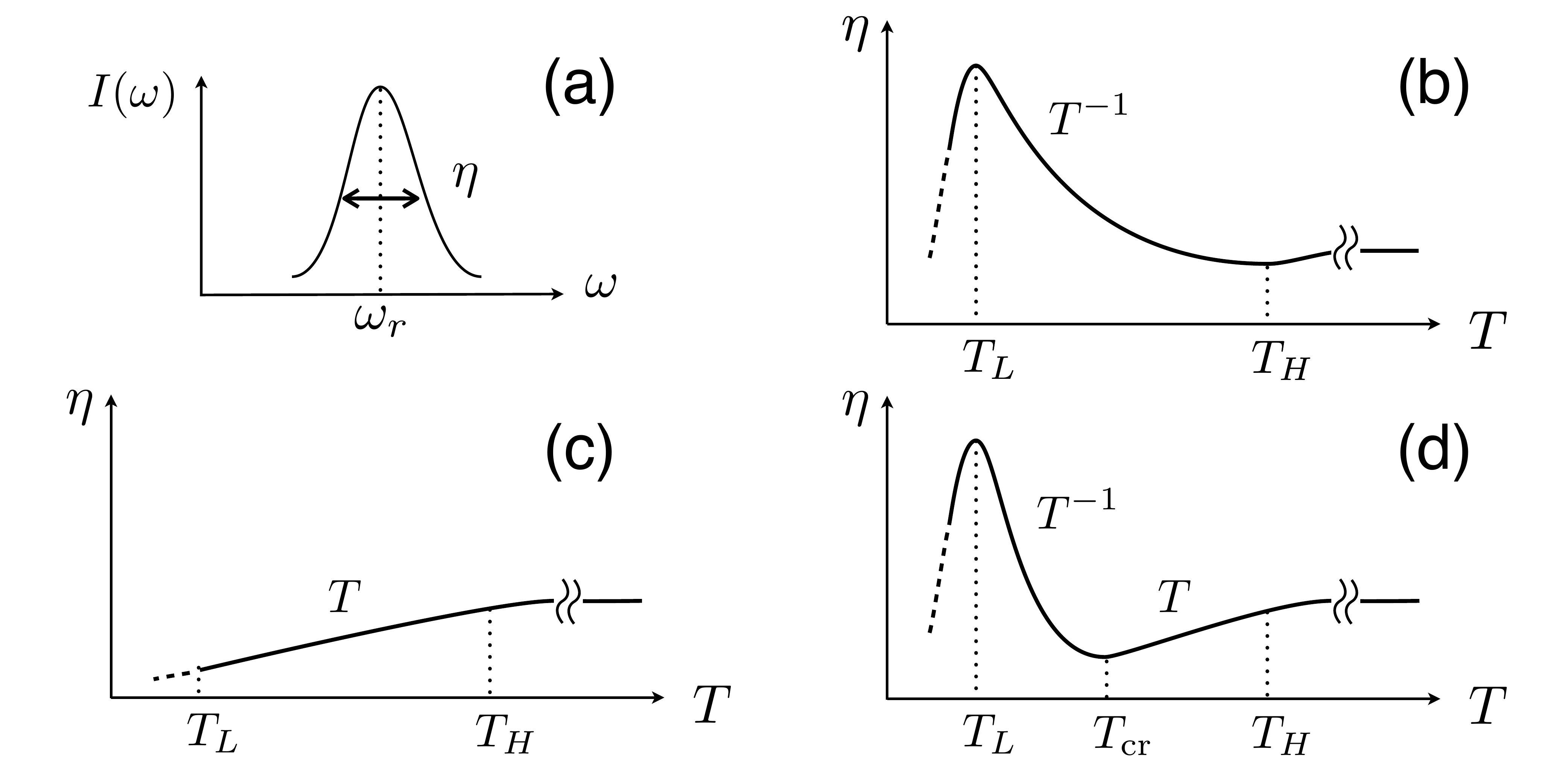}
 \caption{
 (a) The Lorentzian ESR absorption spectrum
 $I(\omega)\propto1/((\omega-\omega_r)^2+\eta^2)$. 
 The other panels show schematic $T$ dependence of the ESR linewidth
 $\eta$ in the strong-leg spin ladder \eqref{eq:H_ladder} with
 (b) the rung anisotropy $\mathcal{H}'_r$, (c) the leg anisotropy
 $\mathcal{H}'_\parallel$ and (c) both of them
 $\mathcal{H}'_\perp+\mathcal{H}'_\parallel$  for $H\ll T$.
 Only in the presence of $\mathcal{H}'_r$, the broadening $\propto
 T^{-1}$ occurs.
 In the high (low) temperature region $T>T_H$ ($T<T_L$), 
 the linewidth is constant (gradually disappears down to
 $T\searrow0$). 
 The similar behavior of $\eta$ is 
 also predicted to be observed in other quasi-1D antiferromagnets (see
 the text and  
 Table~\ref{table:lw}). 
 }
 \label{fig:Tdep}
\end{figure}

\textit{Linewidth of Ladder. ---}
We derive the ESR linewidth of the spin ladder 
in the temperature-induced TLL phase $T_L<T<T_H$. 
At high temperatures $H\ll T$, the linewidth $\eta$ 
[see Fig.~\ref{fig:Tdep}~(a)] is given 
by~\cite{MoriKawasaki,OshikawaAffleck}
\begin{equation}
 \eta
  = -\operatorname{Im} G^R_{\mathcal  A\mathcal A^\dagger}  (\omega =
  H)/2\langle S^z\rangle,
 \label{eq:eta}
\end{equation}
where $G^R_{\mathcal{A}\mathcal{A}^\dagger}(\omega)$ is retarded Green's
function~\cite{suppl} of $\mathcal{A}(t)=[\mathcal{H}',S^+(t)]$ with $S^+=\sum_{j,n}(S^x_{j,n}+iS^y_{j,n})$.
For $\mathcal{H}'=0$, we obtain $\eta=0$.
The ESR absorption peak acquires a nonzero linewidth from the weak
anisotropy $\mathcal{H}'$, namely $\mathcal{A}$.
The temperature-induced TLL has an advantage that Eq.~\eqref{eq:eta} is
analytically computable.

We first consider a longitudinal rung anisotropy 
$\mathcal{H}'=\mathcal{H}'_\perp$,
\begin{equation}
  \mathcal{H}'_\perp=J_\perp\delta_{\perp{}z}\sum_j
   S^z_{j,1}S^z_{j,2}.
\label{eq:H'_r}
\end{equation}
Note that $\delta_{\perp z}$ does not have to be small as far as
$J_\perp/J_\parallel$ is small.
Utilizing the bosonization technique,~\cite{Giamarchi_book} 
we can compute the linewidth driven by the rung 
anisotropy $\mathcal{H}'_\perp$ as~\cite{suppl}
\begin{equation}
 \eta = \eta_\perp
  = \frac{J_\perp^2\delta_{\perp z}^2}{J_\parallel\chi} \biggl(
  \frac{4C_u^2}{\pi} \frac T{J_\parallel} + \frac{\pi^2 C_s^4}{16}
  \frac{J_\parallel}T 
  \biggr).
  \label{eq:eta_rung}
\end{equation}
where $C_{u,s}$ are dimensionless constants that appear in 
bosonization formulas of the spin $\bm{S}_{j,n}$~\cite{Giamarchi_book} 
and $\chi=\langle{S^z_{j,n}}\rangle/H$ is the uniform susceptibility, 
which is insensitive to $T$ in the temperature-induced TLL phase. 
One can similarly treat the effect of a transverse rung
anisotropy $\mathcal{H}'=\mathcal{H}'_\perp=J_\parallel\delta_{\perp
x}\sum_{j}S^x_{j,1}S^x_{j,2}$.
The contribution of the transverse rung
anisotropy to the linewidth is a half of
Eq.~\eqref{eq:eta_rung}.

Equation~\eqref{eq:eta_rung} shows a clear 
difference between WCSCs and purely 1D cases. 
The first term $\propto T/J$ of Eq.~\eqref{eq:eta_rung} comes from a
marginally irrelevant operator in the renormalization group (RG) sense.
Such a marginal interaction appears commonly in quasi-1D
quantum magnets.
In the case of the $S=1/2$ XXZ chain, this $T$-linear term dominates 
the ESR linewidth~\cite{OshikawaAffleck}.
By contrast, the last term $\propto (T/J)^{-1}$ of
Eq.~\eqref{eq:eta_rung} emerges \emph{only after}
the relevant anisotropic interchain interaction is included.
To see this, we next consider an anisotropic exchange interaction on
legs 
$\mathcal{H}'_\parallel=J_\parallel\delta_{\parallel
z}\sum_{j,n}S^z_{j,n}S^z_{j+1,n}$ with $|\delta_{\parallel z}|\ll1$. 
Different from  $\delta_{\perp z(x)}$,
$|\delta_{\parallel z}|$ should be small so that $\mathcal{H}'_\parallel$ is
perturbative.
The anisotropy along the leg $\mathcal{H}'_{\parallel}$
gives the same result as the Oshikawa-Affleck theory for the XXZ
chain,~\cite{OshikawaAffleck} 
$\eta=\eta_\parallel=2J_\parallel\delta_{\parallel
z}^2(4C_u^2/\pi + C_s^4)T/J_\parallel\chi$.
The transverse anisotropy 
$\mathcal{H}'_\parallel=J_\parallel\delta_{\parallel
x}\sum_{j,n}S^x_{j,n}S^x_{j+1,n}$  
gives a linewidth a half of it~\cite{Furuya_lw}.
The direction of anisotropy is irrelevant for the temperature dependence
because it merely modifies a numerical factor of the linewidth.

Spin ladder compounds are generally expected to possess both rung and leg 
anisotropies, $\mathcal{H}'=\mathcal{H}'_\perp+\mathcal{H}'_\parallel$, 
in which the linewidth $\eta$ has two possible dominant terms. 
One comes from the rung anisotropy 
$\eta_\perp\propto J_\perp^2\delta_{\perp z}^2T^{-1}$ and 
the other from the leg anisotropy $\eta_l\propto\delta_{\parallel z}^2T$. 
Given comparable $\delta_{\perp z}$ and $\delta_{\parallel z}$,
the linewidth changes its temperature dependence at 
a crossover temperature $T_{\mathrm{cr}}\sim 
|\delta_{\perp z} J_\perp/\delta_{\parallel z}|$. 
For $T_{\mathrm{cr}}<T<T_H$, the leg anisotropy dominates the linewidth 
and leads to a $T$-linear narrowing. 
For $T_L<T<T_{\mathrm{cr}}$, the rung anisotropy dominates
the linewidth and leads to the $T^{-1}$ broadening
[Fig.~\ref{fig:Tdep}~(d)]. 
We summarized the characteristic temperatures $T_{H,L,\mathrm{cr}}$ in 
Table.~\ref{table:lw}.

\begin{table}[t!]
\begin{tabular}[t]{lcccc}
 \hline
 system & $\eta$ & $T_L$ & $T_{\mathrm{cr}}$ & $T_H$   \\
 \hline
ladder with $\mathcal{H}'_\perp$                  &   $AT^{-1}$    &
     $\Delta$ &         $J_\parallel$        & $J_\parallel$        \\ 
ladder with $\mathcal{H}'_\parallel$   &   $BT$    &
     $\Delta$ &         N/A        &   $J_\parallel$      \\ 
ladder with $\mathcal{H}'_\perp+\mathcal{H}'_\parallel$   &
     $AT^{-1}+BT$    & 
     $\Delta$ &         $|J_\perp|$   &   $J_\parallel$      \\ 
 3D coupled chains  &   $AT^{-1}+BT$    &
     $T_c$   &    $|J_\perp|$ &  $J_\parallel$    \\ 
  \hline
 XXZ chain~\cite{OshikawaAffleck}         &   $BT$         &
	$0$             & $J_\parallel$   &   $J_\parallel$  \\
 a staggered DM~\cite{OshikawaAffleck}&   $C(DH/T)^2$  &
	$(DH)^{2/3}$       &
	     $J_\parallel$  &    $J_\parallel$  \\ 
 \hline
\end{tabular}
 \caption{Orders of characteristic temperatures $T_{L,H,{\mathrm{cr}}}$ 
 and the temperature
 dependence of the ESR linewidth $\eta$ in the (multi-component) TLL
 phase $T_L<T<T_H$ for several $S=1/2$ WCSC.
 Constants $A$, $B$ and $C$ depend on 
 microscopic information on the systems.
 For comparison, we list the results of the intrachain anisotropies,
 XXZ anisotropy and the staggered DM interaction.~\cite{OshikawaAffleck}
 The const. $D$ is the magnitude of the DM vector.}
 \label{table:lw}
\end{table}

One can systematically understand the temperature dependence of the
linewidth from the RG viewpoint. In TLL phases, 
every anisotropic interaction is written as an operator with a scaling
dimension $d_s$ whose coupling constant $\delta$ grows or vanishes 
following the RG equation~\cite{Cardy_RG}. 
The RG argument shows that in the anisotropy term 
$\mathcal{H}'$, (non-chiral) operators with dimension $d_s$ contribute 
to the linewidth in a form
\begin{equation}
 \eta \propto J_\parallel\delta^2(T/J_\parallel)^{2d_s-3}
  \label{eq:eta_pow}
\end{equation}
Equation~\eqref{eq:eta_pow} is derived from the relation 
$\eta\propto\operatorname{Im}G^R_{(d_s/2,d_s/2)}(H,H/v)
\propto(T/J_\parallel)^{2(d_s-2)}\sinh(H/2T)$,
where the Green's function $G^R_{(d_s/2,d_s/2)}(\omega,q)$ is 
given in Eq. (S22) of the Supplementary material~\cite{suppl}. 
Since the power $2d_s-3$ has the one-to-one correspondence with $d_s$,
the power law \eqref{eq:eta_pow} of the linewidth allows us to
determine the most relevant anisotropic interaction in TLL phases of 
WCSC compounds. 
The bosonization theory~\cite{Giamarchi_book} tells us 
that $d_s$ is limited to two values (i) $d_s=1/2$ and (ii)
$d_s=1$ when the magnetic field $H$ is low enough and 
the system is nearly SU(2) symmetric. 
The case (i) corresponds to a case of a staggered field $h$ along the
chain generated from the staggered DM interaction, leading
to~\cite{OshikawaAffleck} $\eta\propto
J(h/J_\parallel)^2(T/J_\parallel)^{-2}$.  
Although the bond alternation also generates an operator with $d_s=1/2$,
it preserves the $SU(2)$ rotational symmetry and has no impact on the
linewidth.
The case (ii) leads to the ``$T^{-1}$ law'' of the linewidth
\eqref{eq:eta_rung} driven by the rung anisotropy.
This is a unique and, simultaneously,
universal feature of interchain interactions.
In fact, such an anisotropic interaction with the scaling dimension $d_s=1$ 
is hardly found in intrachain interactions at low magnetic
field~\cite{Giamarchi_book}. 
The $T^{-1}$ law of the linewidth gives an evidence of the interchain
anisotropy $\mathcal{H}'_\perp$.
It will be practically possible to identify the anisotropic interaction
from the power law \eqref{eq:eta_pow} of the linewidth because only
two well discriminable values $d_s=1/2$ and $d_s=1$ are allowed.

\textit{Generalizations. ---}
Our theory on the two-leg ladder is readily extended to 
the temperature-induced TLL phase of the general 
$N_{\mathrm{leg}}$-leg WCSCs. 
As far as the interchain interaction connects two chains 
(not three or more chains simultaneously), 
the $T$ dependence of the ESR linewidth in spin ladders 
(Fig.~\ref{fig:Tdep} and Table~\ref{table:lw}) are still 
valid even for general WCSC irrespective of geometric 
patterns of interchain (rung) couplings. 

As we already mentioned, one can also extend our theory 
to 3D coupled spin chains. 
Thus we confirmed the broadening of the ESR spectrum near
the 3D ordering temperature, $T\searrow T_c$, 
{\it without making ad hoc assumptions}. 
Namely, we gave a reasonable explanation to
origin of the $T^{-1}$ broadening of the ESR linewidth near $T_c$, 
which has been observed in various 3D coupled spin chains. 
The $T$ dependence of the ESR linewidth 
in temperature-induced TLL phases is summarized 
in Fig.~\ref{fig:Tdep} and Table~\ref{table:lw}.

We note that the growth of the linewidth with lowering temperature is
similar to that of NMR relaxation rate~\cite{Chitra_ladder,
Giamarchi_CoupledLadders,Sato_NMR}.
However, the growth of the ESR linewidth does \emph{not} occur without
any relevant anisotropic interaction \eqref{eq:H'_r}, while
that of the NMR relaxation rate always occurs 
regardless of the existence of anisotropies.

\begin{figure}[t!]
 \centering
 \includegraphics[bb= 0 0 864 504, width=\linewidth]{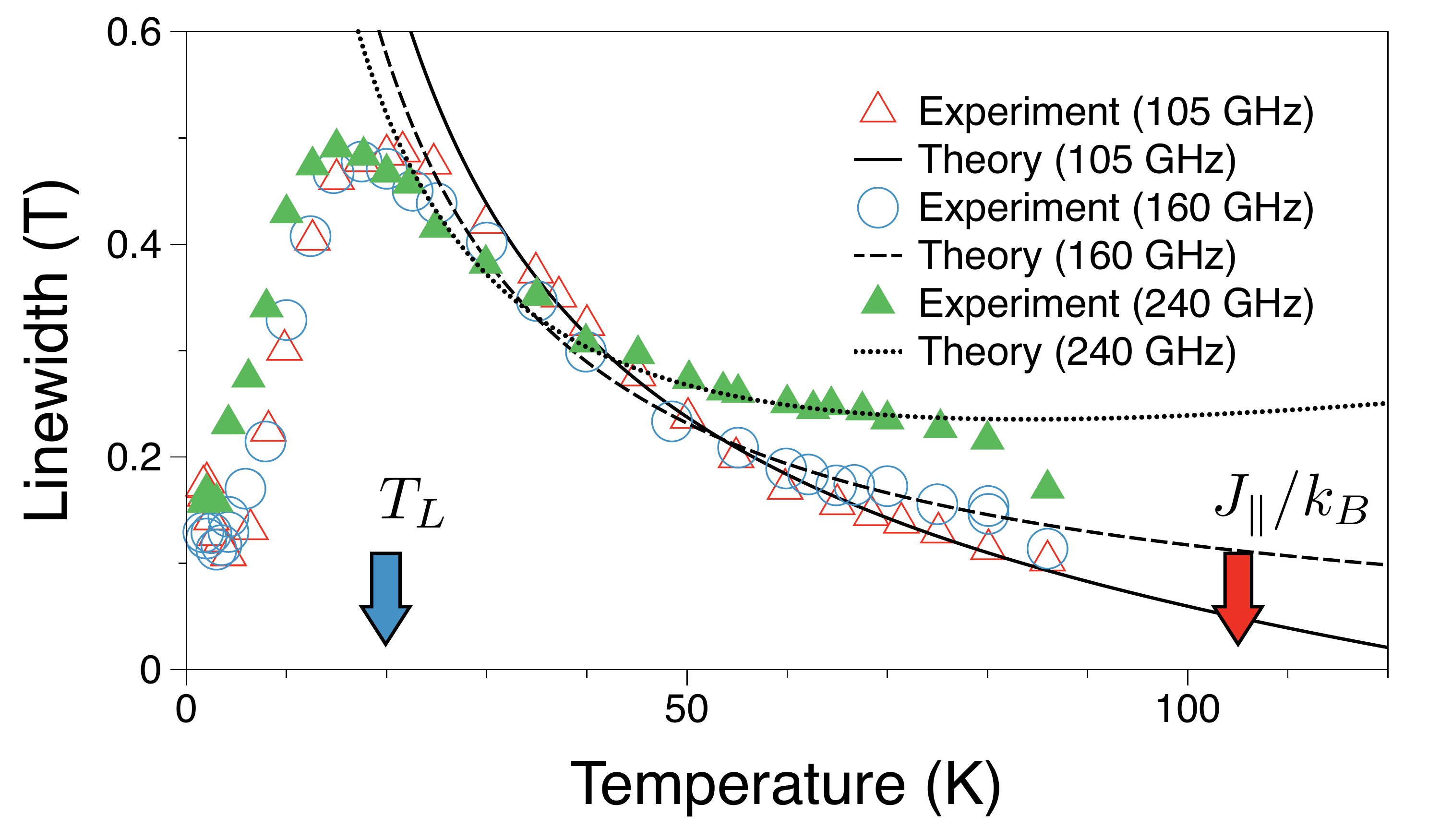}
 \caption{
 Observed ESR linewidth of
 \sul~\cite{Fujisawa_ESR} is compared 
 with our result, $\eta=A/T+BT$ with tuning parameters $A$ and $B$. 
 The exchange narrowing is not found, whcih indicates
 $T_{\mathrm{cr}}\sim J_\parallel$.
 }
 \label{fig:sul}
\end{figure}

\textit{Comparison with Experiments. ---}
Let us apply our results, Fig.~\ref{fig:Tdep} and Table~\ref{table:lw},
to two experiments. 
First we consider the $S=1/2$ four-leg spin tube compound
\sul. Microscopic parameters of the model for this compound is yet to be
be settled~\cite{Zheludev_sul,Shyiko_zigzag}.
Nevertheless, since the $T$ dependence of the susceptibility
$\chi$ for $50$~K $<T<100$~K
is well reproduced with a model of a slightly-bond-alternating chain 
$\mathcal{H}=J_\parallel\sum_{j}(\bm{S}_{2j-1}\cdot \bm{S}_{2j}+
\alpha \bm{S}_{2j}\cdot\bm{S}_{2j+1})$ with $J_\parallel=105.6$~K and 
$\alpha=0.98$~\cite{Fujisawa_JPSJ}, we may conclude that the intrachain
coupling $J_\parallel$ is much stronger than the other interchain
couplings. 
Points of Fig.~\ref{fig:sul} show the experimentally derived 
linewidth~\cite{Fujisawa_ESR}. The set of points for each frequency 
is excellently fitted by a black curve $\eta=A/T+BT$ 
by tuning parameters $A$ and $B$. 
Moreover, the temperature $T_L\sim20$~K, where the linewidth 
takes a maximum value, is almost \emph{independent} of the resonance
frequency $\omega$, namely, the applied magnetic field $H$. 
This behavior eliminates the possibility of the staggered DM interaction
(Table~\ref{table:lw}). 
Besides, we fitted the experimental data with the $d_s=1/2$ formula, that
is, $\eta'=A'/T^2+B'T$ with fitting parameters $A'$ and $B'$.
The linewidth $\eta'$ gives worse fittings of the experimental
data than $\eta=A/T+BT$, especially for $\omega/2\pi=240$~GHz. 
We thus conclude that the DM interaction is unlikely to be the source
of the maximum of the linewidth in \sul.
In the low temperature regime $T<T_L$, a gapped paramagnetic phase is 
expected to be realized due to the interchain interactions (the small bond 
alternation would not be the main origin of the spin gap). 

\begin{figure}[t!]
 \centering
 \includegraphics[bb=0 0 864 504, width=\linewidth]{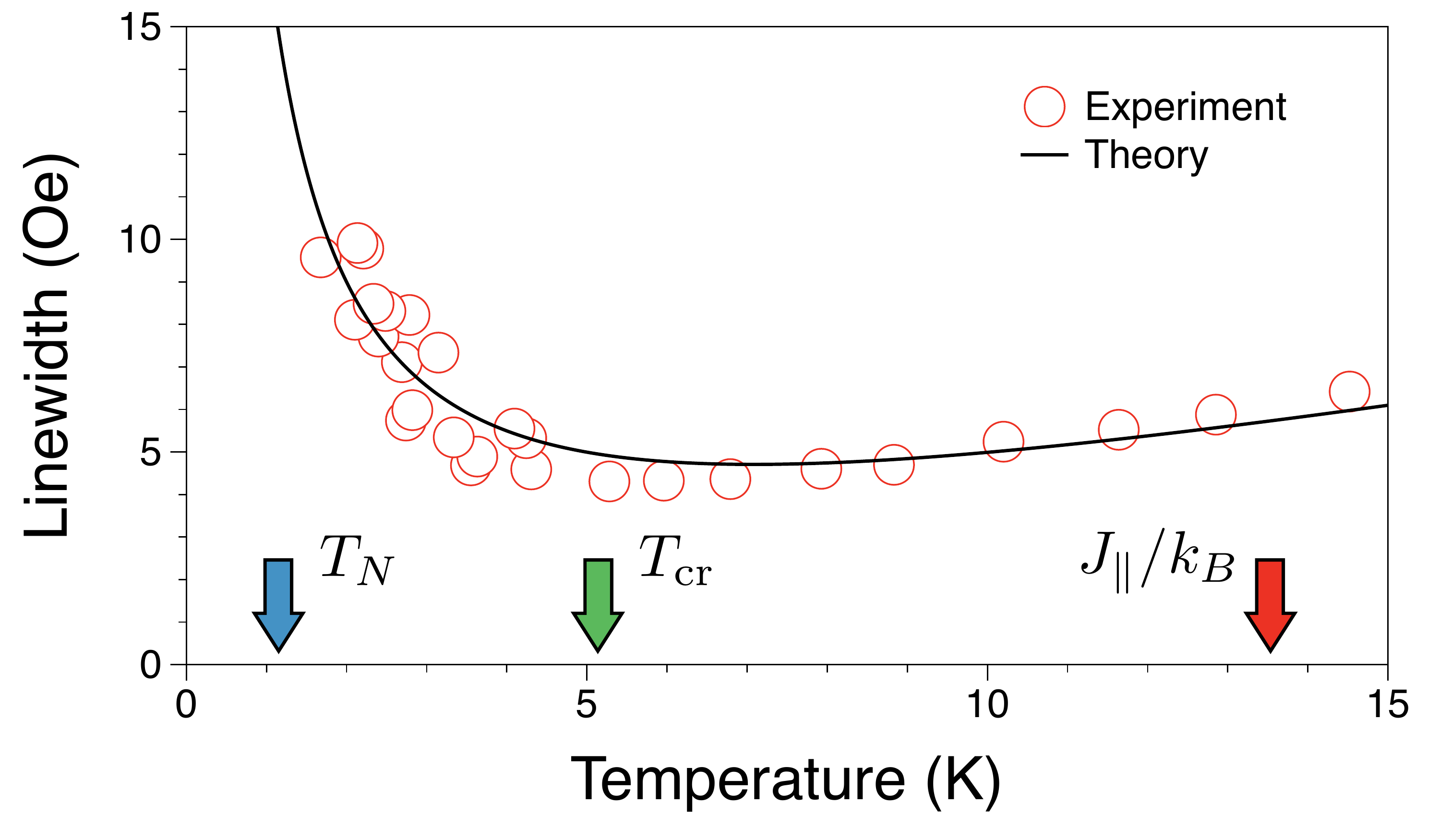}
 \caption{
 ESR linewidth on the $S=1/2$ antiferromagnetic chain compound
 CuCl$_2\cdot2$NC$_5$H$_5$~\cite{Ajiro_ESR} is fitted by using a function
 $\eta=A/T+BT$ with parameters $A$ and $B$.
 }
 \label{fig:3D}
\end{figure}

These results strongly indicate that the region $20$~K $<T<$ $105$~K of 
Figure~\ref{fig:sul} is the temperature-induced TLL phase and that
$T_L\simeq 20$~K will be determined by weak interchain interactions. 
The small ratio $T_L/T_H \sim 10^{-1}$ can be attributed to 
either a case that the interchain interaction is actually about
$10$~\% of the intrachain interaction or a case that the frustrated
interchain interaction lowers the ratio.
From the lattice model proposed by Garlea \textit{et
al.}~\cite{Garlea_sul_PRL}, we speculate that the latter 
possibility is likely to occur. 
Figure~\ref{fig:sul} also tells us that the intrachain anisotropy 
$\mathcal{H}'_\parallel$ is much weaker than
the \emph{interchain} one $\mathcal{H}'_\perp$ because the linewidth
behaves like Fig.~\ref{fig:Tdep}~(b).

Next we consider the $S=1/2$ antiferromagnetic spin chain compound 
CuCl$_2\cdot2$NC$_5$H$_5$. 
The intrachain and 3D interchain couplings are evaluated as 
$J_\parallel=13.4$~K and $J_\perp=0.12$~K, respectively. 
The N\'eel ordered phase appears at $T_N=1.135$~K due to $J_\perp$. 
Figure~\ref{fig:3D} shows the experimental ESR linewidth 
of this compound~\cite{Ajiro_ESR}. The data are well fitted 
by a function $\eta=A/T+BT$ via fitting parameters $A$ and $B$. 
From the black curve, we find a crossover temperature 
$T_{\mathrm{cr}}\simeq 5$~K, which means that intrachain and 
interchain interactions have comparable anisotropies
[Fig.~\ref{fig:Tdep}~(d)].
Since nonuniversal factors $C_{u,s}$ in Eq.~\eqref{eq:eta_rung}
are numerically determined for the spin 
chain~\cite{Lukyanov_amp_PRB, Lukyanov_amp_NPB, 
Hikihara_amp,Takayoshi_amp}, further experiments with changing the
direction of the magnetic field will enable us to determine the strength of
anisotropic exchange interactions of CuCl$_2\cdot2$NC$_5$H$_5$.

\textit{Conclusions. ---}
In this Paper, we presented the organized theory of the
universal features of the ESR linewidth in quasi-1D spin systems.
We uncovered a microscopic origin of the observed broadenings
in the four-leg spin tube compound (Fig.~\ref{fig:sul}) and in
the quasi-1D compound (Fig.~\ref{fig:3D}) in the unified way. 
We emphasize that our theory is applicable to various spatially
anisotropic quantum antiferromagnets even when the low-temperature phase
is nonmagnetic.
Combining our theory on the linewidth with analyses on the resonance
frequency~\cite{Furuya_ladder, Furuya_lw} will allow us to determine
microscopic anisotropic interactions of WCSC schematically.
It will be also interesting to apply our theory to
spatially anisotropic frustrated quantum antiferromagnets such as
Cs$_2$CuCr$_4$~\cite{Kohno_Cs2CuCr4,Starykh_Cs2CuCr4}.

The authors are grateful to Masashi Fujisawa for providing us the
experimental data on \sul{} and for fruitful discussions from the
experimental viewpoint.
We also thank Thierry Giamarchi, Masaki Oshikawa and Akiyuki Tokuno for
useful comments. 
S.C.F. was supported by the Swiss National Foundation under
MaNEP and division II. M. S. was supported by KAKENHI (Grant
No. 25287088 and No. 26870559).

\section*{Supplemental Material}

\section{Bosonization for Ladders}
In this section, we describe technical details of non-Abelian and
Abelian bosonizations for the $S=1/2$ two-leg spin ladder of Eq.~(1). 
Suppose the interchain (rung) coupling is sufficiently weaker than the 
intrachain (leg) coupling, then we may start with two decoupled 
$S=1/2$ Heisenberg Antiferromagnetic (AF) chains to analyze the ladder. 
The low-energy physics of a $S=1/2$ single AF chain $\mathcal{H}_n$ 
is described by a gapless Tomonaga-Luttinger liquid (TLL). 
In the bosonization framework, the spin operator $\bm S_{j,n}$ on 
the $n$th leg is written in two parts:~\cite{Giamarchi_book,
Tsvelik_book, AffleckHaldane} 
\begin{equation}
 \bm S_{j,n} = \bm J_n + (-1)^j \bm N_n.
  \label{eq.S2JN}
\end{equation}
The uniform component $\bm J_n = \bm J_{nL} + \bm J_{nR}$ is split into
left-moving ($\bm J_{nL}$) and right-moving ($\bm J_{nR}$) parts.
These fields are written in terms of chiral bosons $\phi_{nL(nR)}$:
\begin{align}
 J^z_{nL} 
 & = \frac M2 + \frac 1{\sqrt{2\pi}}\partial_x\phi_{nL}, \\
 J^z_{nR}
 & = \frac M2 + \frac 1{\sqrt{2\pi}}\partial_x\phi_{nR}, \\
 J^\pm_{nL}
 & = \frac{C_u}2 e^{\pm i (\sqrt{8\pi}\phi_{nL}+Hx/v)}, \\
 J^\pm_{nR}
 & = \frac{C_u}2 e^{\mp i( \sqrt{8\pi}\phi_{nR} + Hx/v)},
\end{align}
where $M$ is the longitudinal magnetization per a site
induced by a magnetic field $H$, $C_u$ is a 
nonuniversal dimensionless constant, and $x=ja$ is the continuous
coordinate ($a$ is the lattice constant). 
The boson field $\phi_n$ and its dual field
$\theta_n$ are respectively defined as 
\begin{equation}
 \phi_n  = \phi_{nL} + \phi_{nR}, \qquad
 \theta_n   = \phi_{nL} - \phi_{nR}.
\end{equation}
The TLL Hamiltonian is expressed by using these fields as follows: 
\begin{equation}
\mathcal{H}_n\simeq \int dx\, \frac{v}{2}
[K(\partial_x\theta_n)^2+K^{-1}(\partial_x\phi_n)^2],
\end{equation}
where $v$ is the velocity of the TLL and $K$ is the TLL parameter. 
If the chain $\mathcal{H}_n$ has the spin-rotational SU(2) symmetry, 
the value of $K$ is fixed to unity, while magnetic anisotropies and 
external field $H$ generally change the value. 
The staggered component $\bm N_n$ is written by using an SU(2) matrix field
$\mathcal U_n$:
\begin{equation}
 \bm N_n = C_s \Tr (\mathcal U_n \bm \sigma),
\end{equation}
where $C_s$ is a dimensionaless constant and 
$\bm \sigma = (\sigma^x, \sigma^y, \sigma^z)$ is a set of Pauli
matrices,
\begin{equation}
 \sigma^x
  =
  \begin{pmatrix}
   0 & 1 \\
   1 & 0
  \end{pmatrix}, \quad
 \sigma^y
  =
  \begin{pmatrix}
   0 & -i \\
   i & 0
  \end{pmatrix}, \quad
  \sigma^z
  =
  \begin{pmatrix}
   1 & 0 \\
   0 & -1
  \end{pmatrix}.
\end{equation}
The element of the matrix field $\mathcal U_n$ can be represented as
\begin{equation}
 \mathcal U_n
  = \frac{-i}2
  \begin{pmatrix}
   e^{i\sqrt{2\pi}\phi_n} & ie^{-i\sqrt{2\pi}\theta_n} \\
   ie^{i\sqrt{2\pi}\theta_n} & e^{-i\sqrt{2\pi}\phi_n}
  \end{pmatrix}.
\end{equation}

Let us briefly discuss effects of the rung coupling and 
the excitation gaps of the ladder at zero magnetic field.
Using the bosonization formula \eqref{eq.S2JN}, the rung interaction
(3) in the paper is represented as
\begin{align}
 \hr
 & = g_r \int dx \biggl( \cos \sqrt{4\pi}\theta_-
 \notag \\
 & \quad +
 \frac 12 \cos \sqrt{4\pi}\phi_- + \frac 12
 \cos \sqrt{4\pi}\phi_+\biggr),
 \label{eq.hr_phi}
\end{align}
where $g_r = J_rC_s^2$, and we have introduced new symmetric ($+$) 
and antisymmetric ($-$) fields 
\begin{equation}
\phi_\pm = (\phi_1 \pm \phi_2)/\sqrt{2},\,\,\,\,\,\,
\theta_\pm = (\theta_1 \pm \theta_2)/\sqrt{2}.
\end{equation}
This rung interaction \eqref{eq.hr_phi} generates mass gaps of the 
$\phi_\pm$ fields. 
According to Shelton, Nersesyan and Tsvelik~\cite{Shelton_fermionization}, 
under zero magnetic field $H=0$, two boson fields $\phi_\pm$ and 
their dual fields $\theta_\pm$ can be fermionized and the
fermionized theory contains four Majorana fermions. 
Three of them have a degenerate excitation gap $\Dt = |g_r|/2$ and the 
other one has a larger gap $\Ds = 3|g_r|/2$. 
Physically, $\Dt$ is the spin-triplet (magnon) excitation gap 
and $\Ds$ is the spin-singlet (two-magnon bound state) gap. 
The lowest-energy excitation gap at zero magnetic field
equals to $\Delta_t$. Since the uniform component of spins 
$\bm J_1+\bm J_2$ is proportional to $\partial_x\phi_+$, 
only the symmetric field $\phi_+$ is coupled to an external magnetic 
field $H$ in the Zeeman term $-H\sum_j\bm S_{1,j}+\bm S_{2,j}$.

\section{Retarded Green's functions}
Let us discuss several important points in the practical
calculation of the  electron spin resonance (ESR) spectrum of the spin
ladder~(1).  
We consider the high-temperature region, $T \gtrsim \Delta$, in which 
the rung interaction $\hr$ and the perturbative anisotropy term
$\mathcal{H}'$ are both negligible in calculations of thermodynamic
properties.   
In this region, the two-leg ladders are described by two TLLs:
\begin{equation}
 \mathcal H_{\mathrm{ladder}} \simeq \int dx \sum_{a= \pm }
  \frac v2 \bigl\{ (\partial_x   \theta_a)^2 + (\partial_x
  \phi_a)^2\bigr\}  
\end{equation}
We should however note that, when the ESR spectrum is concerned,
we cannot neglect the effect of the magnetic anisotropy $\mathcal{H}'$ 
even in the high temperature region. 
This is because the ESR spectrum is essentially determined by the
magnetically anisotropic terms even if they are very small.

Here we focus on the following longitudinal rung anisotropy:
\begin{equation}
 \mathcal H' = J_r \delta_{\perp z} \sum_j S_{j,1}^z S_{j,2}^z.
  \label{eq.H'}
\end{equation}
For this anisotropy, the uniform and staggered components of the
operator  $\mathcal A = [\mathcal H', S^+]$, that is,
\begin{equation}
 \mathcal{A} = 2J_r \delta_{\perp{}z} (\mathcal{J}_\perp +
  \mathcal{N}_\perp),
  \label{eq.A_rung}
\end{equation}
where  $\mathcal J_\perp$ and $\mathcal N_\perp$ can be expressed 
in terms of the boson fields as follows: 
\begin{align}
 \mathcal J_\perp
 & =  \int dx \sum_{n\not=n'} \bigl( :J^z_{nL}:
 J^+_{n'R}+:J^z_{nR}:J^+_{n'L}\bigr) \notag \\
 & = \frac{C_u}{\sqrt{4\pi}} \int dx \,\bigl[
 \partial_x
 \phi_{-L} e^{-i\sqrt{4\pi}\phi_{+R}-iHx/v} \cos \sqrt{4\pi}\phi_{-R}
 \notag \\
 & \qquad  +\partial_x \phi_{+R}
  e^{i\sqrt{4\pi}\phi_{+L} + iHx/v} \cos \sqrt{4\pi}\phi_{-L}
  \bigr] \notag \\
  & \quad +i\frac{C_u}{\sqrt{4\pi}} \int dx \, \bigl[ 
  \partial_x\phi_{-L} e^{-i\sqrt{4\pi}\phi_{+R}-iHx/v} \sin
  \sqrt{4\pi}\phi_{-R} \notag \\
 & \qquad- \partial_x \phi_{-R} e^{i\sqrt{4\pi}\phi_{+L}
 +iHx/v} \sin \sqrt{4\pi}\phi_{-L}\bigr],
 \label{eq.J}
 \end{align} 
and
\begin{align}
 \mathcal N_\perp
 & = \int dx \, (N^z_1 N^+_2 + N^z_2 N^+_1) \notag \\
 & = C_s^2 \int dx \, \bigl[ \sin (\sqrt{4\pi}\phi_{+L}+Hx/v)
 e^{-i\sqrt{4\pi}\phi_{-R}} \notag \\
 & \qquad + \sin (\sqrt{4\pi}\phi_{+R} + Hx/v)
 e^{i\sqrt{4\pi}\phi_{-L}}\bigr]. 
 \label{eq.N}
\end{align}
$:\mathcal{O}:=\mathcal{O}-\langle\mathcal{O}\rangle$ 
denotes a normal ordering. 
Here chiral boson fields $\phi_{\pm L(R)}$ have been defined by
\begin{equation}
 \phi_{\pm} = \phi_{\pm L} + \phi_{\pm R}, \qquad
  \theta_{\pm} = \phi_{\pm L} - \phi_{\pm R}.
\end{equation}
Following the standard bosonization technique, 
the Fourier transformed retarded Green's function of the operator
$\mathcal J_\perp$ is calculated as 
\begin{align}
 G^R_{\mathcal J_\perp \mathcal J_\perp^\dagger}(\omega)
 & = \frac{NC_u^2}{\pi}G^R_{(1,1)}(\omega, \tfrac Hv),
\end{align}
where $N$ is the length of the leg and the Green's function 
$G^R_{(1,1)}(\omega, \tfrac Hv)$ is defined as
\begin{equation}
 G^R_{(\frac d2,\frac d2)} (\omega,q)
 = -\sin(\pi d)
 F_d\biggl(\frac{\omega- vq}{4\pi T}\biggr)
 F_d\biggl(\frac{\omega+vq}{4\pi T}\biggr),
 \label{eq.GR_pre}
\end{equation}
with $d=2$.
Here $F_x(y)=(2\pi T/v)^{x-1}B(\frac{x}2-iy,1-x)$ with
$B(x,y)$ being the Beta function. 
The quantity $G^R_{\mathcal{A}\mathcal{A}^\dagger}(\omega=H)$ in
Eq.~(6) is written in terms of 
$G^R_{(\Delta,\bar\Delta)}(\omega,q)$ as
\begin{equation}
 G^R_{\mathcal{A}\mathcal{A}^\dagger}(H)
  = 4J_r^2\delta_{\perp z}^2
  \biggl[ \frac{C_u^2}{\pi}G^R_{(1,1)}(H,\tfrac Hv) +
 \frac{C_s^4}4G^R_{(\frac  12,\frac  12)}(H,\tfrac Hv)\biggr],
 \label{eq.GR_perp}
\end{equation} 

In order to obtain the ESR spectrum, we need 
$G^R_{\mathcal J_\perp \mathcal J_\perp^\dagger}(\omega=H)$, namely, 
$G^R_{(1,1)}(H, H/v)$.
Instead of the general representation for 
$G^R_{(\frac d2, \frac d2)}(\omega, q)$ of Eq.~\eqref{eq.GR_pre}, we
here comment on another useful representation,
\begin{align}
 & G^R_{(\frac d2, \frac d2)}(\omega, q)\notag \\
 & =  -\frac 1{\sin(\pi d)} \frac 1{\Gamma^2(d)} \biggl(
 \frac{2\pi T}v\biggr)^{2(d-1)} \notag \\
 & \quad \times \biggl| \Gamma\biggl( \frac d2 + i\frac{\omega -
 vq}{4\pi T}\biggr)\Gamma\biggl( \frac d2 + i\frac{\omega + vq}{4\pi
 T}\biggr)\biggr|^2 \notag \\
 & \quad \times \sin \biggl( \frac{\pi d}2 + i\frac{\omega - vq}{4\pi
 T}\biggr)\sin\biggl( \frac{\pi d}2 + i \frac{\omega + vq}{4\pi
 T}\biggr), 
 \label{eq.GR}
\end{align}
where $\Gamma(z)$ is the Gamma function.
The equivalence between Eq.~\eqref{eq.GR_pre} and Eq.~\eqref{eq.GR} is
easily proven by using a definition of the Beta function,
$B(x,y) = \Gamma(x)\Gamma(y)/\Gamma(x+y)$
and a reflection formula of the Gamma function,
\begin{equation}
 \Gamma(z) \Gamma(1-z) = \frac{\pi}{\sin(\pi z)}.
\end{equation}
Taking the limit $d \to 2$ in Eq.~\eqref{eq.GR}, we obtain
\begin{align}
 G^R_{(1,1)}(H,  H/v)
 & = \lim_{d \to 2} \frac{\sin (\pi d/2)}{\sin(\pi d)} \biggl(
 \frac{2\pi T}v\biggr)^2 \notag \\
 & \qquad \times \biggl| \Gamma\biggl(1+ i\frac H{2\pi
 T}\biggr)\biggr|^2  
  \sin \biggl( \frac{iH}{2 T}\biggr) \notag \\
 & \simeq -i\frac{\pi^2 HT}{v^2}.
\end{align}
It is purely imaginary.

On the other hand, $G_{\mathcal N_\perp \mathcal
N_\perp^\dagger}(\omega)$ is proportional 
to $G^R_{(\frac 12, \frac 12)}(\omega, H/v)$:
\begin{equation}
 G^R_{\mathcal N_\perp \mathcal N_\perp^\dagger}(\omega)
   = \frac{NC_s^4}4 G^R_{(\frac 12, \frac 12)}(\omega, \tfrac Hv).
\end{equation}
The right hand side is divergent if we take $\omega = H$.
In fact, if we put $d = 1+\Delta'$ with $|\Delta'| \ll 1$ 
in Eq.~\eqref{eq.GR}, we obtain
\begin{align}
 G^R_{(\frac d2, \frac d2)} (H, \tfrac Hv)
 & = - \frac 1{\sin (\pi d)}\Gamma^4(\tfrac 12) \sin
 \biggl( \frac{\pi d}2 + i \frac H{2 T}\biggr) 
 \notag \\
 & \simeq \frac{\pi}{\Delta'} -i \frac{\pi^2H}{4T},
 \label{eq.GR_1/2}
\end{align}
where we have considered the case $H \ll T$. 
As $\Delta'$ approaches zero, the real part diverges.
However, we can show that this divergence of the real part has 
no impact on observable quantities due to the following reasons.
First we should note that an infinitesimal $\Delta'$ is actually
nonzero if we include the renormalization of the TLL parameter 
in the spin chain $\mathcal H_n$. 
For an SU(2)-symmetric AF spin chain, the TLL parameter is fixed, 
but any magnetic anisotropy generally change the value of the TLL
parameter  and it also induces the change of scaling dimensions $d$ of 
vertex operators $e^{i\alpha\theta_\pm}$ and
$e^{i\alpha\phi_\pm}$~\cite{Giamarchi_book}.  
After including the renormalization due to the anisotropy of
Eq~\eqref{eq.H'},  
$\Delta'$ is approximately given by
\begin{equation}
 \Delta' \simeq \frac 12 \biggl( \frac{J_r\delta_{\perp z}}{2\pi^2
  J}\biggr)^2. 
\end{equation}
Since $\mathcal A=[\mathcal H',S^+]$ is proportional 
to the small parameter $\delta_{\perp}$, 
we did not need to carefully consider the effect of the renormalized 
TLL parameter on $\mathcal J_\perp$ and $\mathcal N_\perp$ in the main text. 
Namely, the renormalization effect leads to just a higher order 
correction to $\mathcal A$ and the ESR spectrum. 
Second we find that the divergent term $\propto{\Delta'}^{-1}$ 
in Eq.~\eqref{eq.GR_1/2} cancels out the small factor $\delta_{\perp
z}^2$ in Eq.~\eqref{eq.GR_perp}. As a whole, the physical
quantity  
$G^R_{\mathcal A\mathcal A^\dagger}(H)$ hence becomes \emph{finite}. 
The point is that the divergent correlation 
$G^R_{(\frac 12, \frac 12)}(\omega)$ always appears in physical quantities 
with a form $\delta_{\perp z}^2 G^R_{(\frac 12, \frac 12)}(\omega)$ 
so as to be finite.

\end{document}